\documentclass{ws-ijmpd}
\usepackage[super,compress]{cite}
\usepackage{hyperref}
\usepackage{url}
\usepackage{graphicx}
\usepackage{color}
\usepackage[lflt]{floatflt}

\begin{document}

\newcommand{\changeR}[1]{\textcolor{blue}{#1}}
\newcommand{\changeRII}[1]{\textcolor{red}{#1}}
\newcommand{\DTe}{{{\Delta_{T_e}}}}
\newcommand{\SQ}{{{\mathcal{E}}}}
\newcommand{\TBB}{{{T_{\rm BB}}}}
\newcommand{\TBE}{{{T_{\rm BE}}}}
\newcommand{\TCMB}{{{T_{\rm CMB}}}}
\newcommand{\Te}{{{T_{\rm e}}}}
\newcommand{\Teq}{{{T^{\rm eq}_{\rm e}}}}
\newcommand{\Ti}{{{T_{\rm i}}}}
\newcommand{\nB}{{{n_{\rm B}}}}
\newcommand{\nHe}{{{n_{\rm ^4He}}}}
\newcommand{\nH}{{{n_{\rm H}}}}
\newcommand{\nHet}{{{n_{\rm ^3He}}}}
\newcommand{\nHt}{{{n_{\rm { }^3H}}}}
\newcommand{\nHtw}{{{n_{\rm { }^2H}}}}
\newcommand{\nBes}{{{n_{\rm { }^7Be}}}}
\newcommand{\nBE}{{{n_{\rm BE}}}}
\newcommand{\Bes}{{{{\rm { }^7Be}}}}
\newcommand{\nLis}{{{n_{\rm { }^7Li}}}}
\newcommand{\nLisi}{{{n_{\rm { }^6Li}}}}
\newcommand{\nS}{{{n_{\rm S}}}}
\newcommand{\nSS}{{{n_{\rm ss}}}}
\newcommand{\Teff}{{{T_{\rm eff}}}}
\newcommand{\Mpc}{{{{\rm ~Mpc}}}}

\newcommand{\id}{{{\rm d}}}
\newcommand{\aR}{{{a_{\rm R}}}}
\newcommand{\bR}{{{b_{\rm R}}}}
\newcommand{\neb}{{{n_{\rm eb}}}}
\newcommand{\neql}{{{n_{\rm eq}}}}
\newcommand{\kB}{{{k_{\rm B}}}}
\newcommand{\EB}{{{E_{\rm B}}}}
\newcommand{\zmin}{{{z_{\rm min}}}}
\newcommand{\zmax}{{{z_{\rm max}}}}
\newcommand{\zinj}{{{z_{\rm inj}}}}
\newcommand{\YBEC}{{{Y_{\rm BEC}}}}
\newcommand{\yg}{{{y_{\rm \gamma}}}}
\newcommand{\y}{{{y}}}
\newcommand{\rhob}{{{\rho_{\rm b}}}}
\newcommand{\Ne}{{{n_{\rm e}}}}
\newcommand{\sigT}{{{\sigma_{\rm T}}}}
\newcommand{\me}{{{m_{\rm e}}}}
\newcommand{\npl}{{{n_{\rm pl}}}}
\newcommand{\nY}{{{n_{\rm Y}}}}

\newcommand{\kD}{{{{k_{\rm D}}}}}
\newcommand{\KC}{{{{K_{\rm C}}}}}
\newcommand{\KdC}{{{{K_{\rm dC}}}}}
\newcommand{\Kbr}{{{{K_{\rm br}}}}}
\newcommand{\zdC}{{{{z_{\rm dC}}}}}
\newcommand{\zbr}{{{{z_{\rm br}}}}}
\newcommand{\aC}{{{{a_{\rm C}}}}}
\newcommand{\adC}{{{{a_{\rm dC}}}}}
\newcommand{\abr}{{{{a_{\rm br}}}}}
\newcommand{\gdC}{{{{g_{\rm dC}}}}}
\newcommand{\gbr}{{{{g_{\rm br}}}}}
\newcommand{\gff}{{{{g_{\rm ff}}}}}
\newcommand{\xe}{{{{x_{\rm e}}}}}
\newcommand{\alphafs}{{{{\alpha_{\rm fs}}}}}
\newcommand{\YHe}{{{{Y_{\rm He}}}}}
\newcommand{\SE}{{{\dot{{Q}}}}}
\newcommand{\SN}{{\dot{{N}}}}
\newcommand{\muc}{{{{\mu_{\rm c}}}}}
\newcommand{\xc}{{{{x_{\rm c}}}}}
\newcommand{\xH}{{{{x_{\rm H}}}}}
\newcommand{\mT}{{{{\mathcal{T}}}}}
\newcommand{\mG}{{{{\mathcal{G}}}}}
\newcommand{\Ob}{{{{\Omega_{\rm b}}}}}
\newcommand{\Or}{{{{\Omega_{\rm r}}}}}
\newcommand{\Odm}{{{{\Omega_{\rm dm}}}}}
\newcommand{\mdm}{{{{m_{\rm WIMP}}}}}

\markboth{Sunyaev and Khatri}
{Unavoidable CMB spectral features and blackbody photosphere of our Universe}

\catchline{}{}{}{}{}

\title{\uppercase{Unavoidable CMB spectral features and blackbody photosphere of our 
Universe}\footnote{Based on a talk presented at the Thirteenth Marcel Grossmann Meeting on General Relativity, Stockholm, July 2012.}}

\author{\uppercase{Rashid A. Sunyaev}}

\address{Max Planck Institut f\"{u}r Astrophysik,
  Karl-Schwarzschild-Str. 1\\ 
  85741, Garching, Germany\\
Space Research Institute, Russian Academy of Sciences, Profsoyuznaya
 84/32\\ 117997 Moscow, Russia\\
Institute for Advanced Study, Einstein Drive, Princeton, New Jersey 08540, USA\\
sunyaev@mpa-garching.mpg.de}

\author{\uppercase{Rishi Khatri}}

\address{Max Planck Institut f\"{u}r Astrophysik, Karl-Schwarzschild-Str. 1\\
  85741, Garching, Germany\\
khatri@mpa-garching.mpg.de}

\maketitle

\begin{history}
\received{February 25, 2013}
\revised{March ??, 2013}
\end{history}

\begin{abstract}
Spectral features in 
the %bob
CMB energy spectrum contain a wealth of information
about the physical processes in the early Universe, $z\lesssim 2\times 10^6$. The
CMB spectral distortions are complementary to all other probes of
cosmology. In fact, most of the information contained in 
the %bob
CMB spectrum is
inaccessible by any other means. This review outlines the main physics
behind the spectral features in 
the %bob
CMB throughout the history of the Universe,
concentrating on the distortions which are inevitable and must be present
at a level observable by the next generation of proposed CMB
experiments. The spectral distortions considered here include spectral features from cosmological
recombination, resonant scattering of CMB  by metals during reionization which
allows us to measure their abundances, $y$-type distortions during and
after reionization and  $\mu$-type and $i$-type (intermediate between $\mu$
and $y$) distortions created at
redshifts $z\gtrsim 1.5\times 10^4$.
\end{abstract}

\keywords{CMB,Cosmology,Origin and formation of the Universe,Background radiations,Observational cosmology,Intergalactic matter}

\ccode{PACS numbers:98.80.-k,98.80.Bp,98.70.Vc,98.80.Es,98.62.Ra  }

%\tableofcontents

\section{Spectral distortions of CMB}	

The remarkable measurement of 
the %bob
cosmic microwave background spectrum
(CMB)
by COBE/FIRAS \cite{cobe} showed that the CMB is almost a perfect
blackbody with temperature $\TCMB=2.725\pm 0.001~{\rm K}$ and was not able
to detect any distortions from 
a %bob
blackbody. However, 
the %bob
standard model of  cosmology predicts
distortions in the spectrum  from processes
%bob no comma here
 which heat, cool,
scatter and create CMB photons, throughout most of the history of the Universe.
Many of these processes are connected with absolutely new physics. The goal
of the present paper is not to list all possible sources of distortions
from physics beyond the standard model but to 
%bob "direct attention to the" or "point out the" 
point out the
unavoidable distortions predicted in the standard model. These distortions
are small but fortunately significant progress in technology in the
last two decades permits   an improvement of 2--3 orders of magnitude over
COBE/FIRAS \cite{fm2002}. New proposals like Pixie \cite{pixie},  CoRE
\cite{core} and LiteBIRD \cite{litebird} promise to detect 
the %bob
majority of the unavoidable spectral
distortions we discuss in this paper. Pixie will be able to make absolute
measurements as well as measure anisotropies with 
an %bob
angular resolution
of %bob
$2.6^{\circ}$. CoRE and LiteBIRD will be able to measure only the frequency
dependent anisotropies with
high sensitivity and have proposed
angular resolutions of $\sim 5'$ and $\sim 30'$ respectively. The detection of these spectral
distortions would provide new information about the important properties
of the Universe such as reionization and formation of  
the %bob
first stars and
galaxies at redshifts $6\lesssim z\lesssim 30$, recombination of hydrogen and helium
at redshifts  $z\sim 1100-6000$ and energy injection at redshifts
$1.5\times 10^4 \lesssim z \lesssim 2\times 10^6$ due to the
  dissipation of sound waves in the primordial plasma. We will also discuss
the effect of rapidly cooling baryons and electrons leading, under some
additional conditions, to Bose-Einstein condensation of CMB photons.

\section{Line features from the epoch of hydrogen and helium recombination}
\begin{figure}
\resizebox{\hsize}{!}{\includegraphics{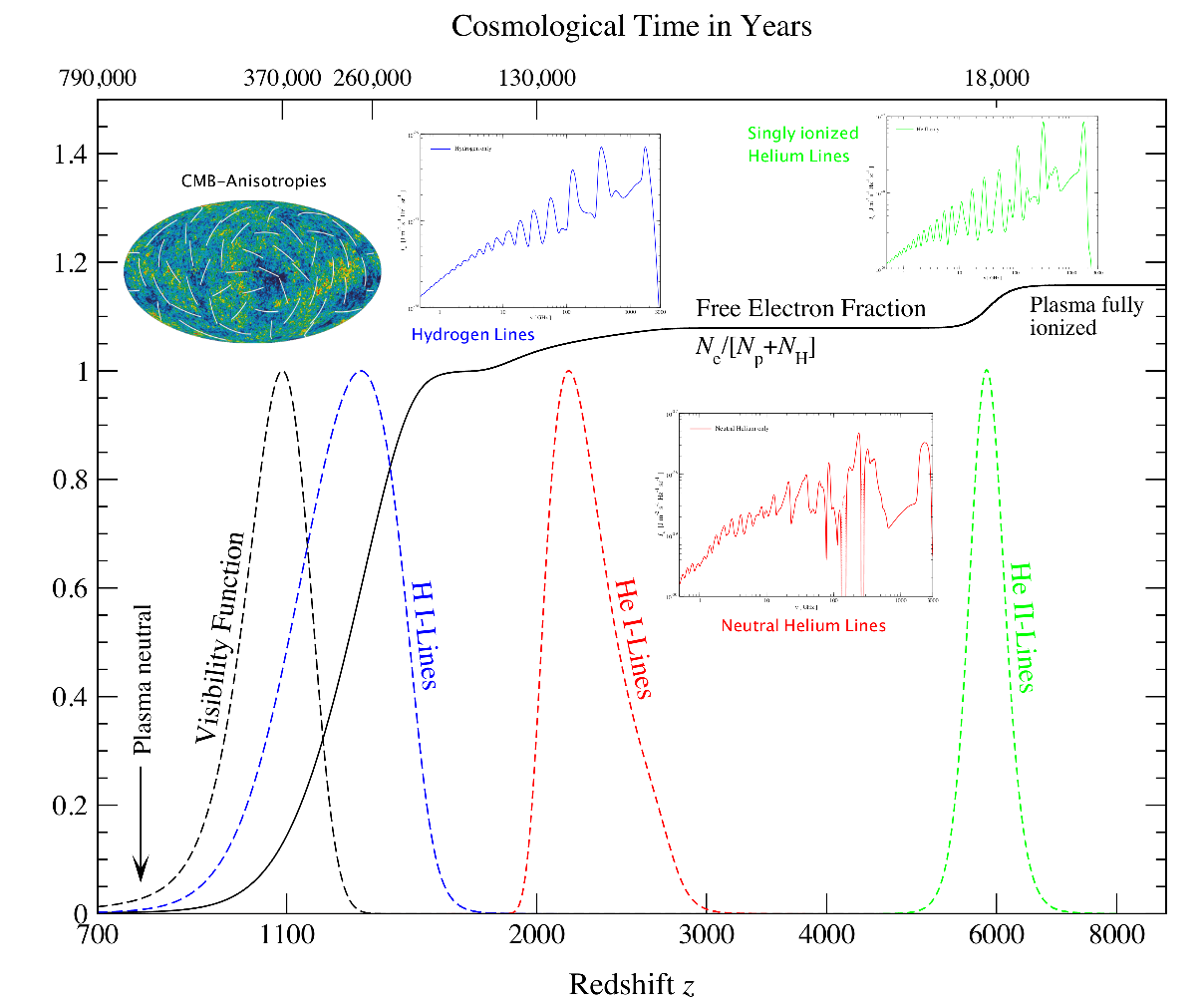}}
\caption{\label{linevisfig} Redshift dependence of the brightness of   the recombination lines from
  the epoch of helium and hydrogen recombination. For comparison, we also
  present the
  visibility function of the last scattering surface corresponding to the
  formation of the CMB acoustic peaks \cite{sz1970c,jw1985,wmap1}. Figure is taken from 
 Ref.~\protect\refcite{sc2009}.}
\end{figure}
\begin{figure}
\resizebox{\hsize}{!}{\includegraphics{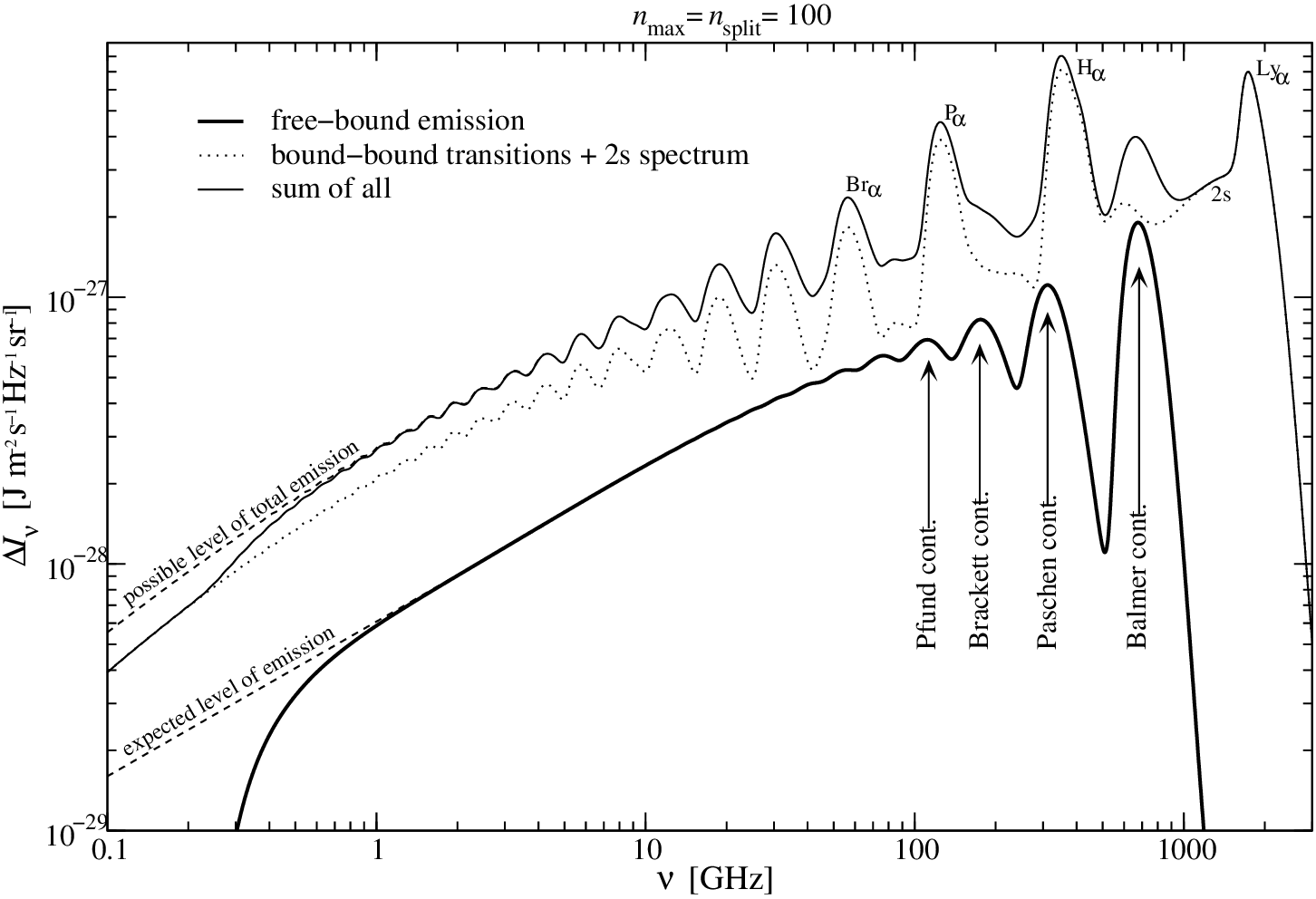}}
\caption{\label{hrecfig} Spectral features resulting from cosmological
  recombination of hydrogen at redshift $z\sim 1300$. For example, optical H$\alpha$ line is
  redshifted to sub-mm spectral band. Figure is taken from 
 Ref.~\protect\refcite{cs2006b}.}
\end{figure}
\begin{figure}
\resizebox{\hsize}{!}{\includegraphics{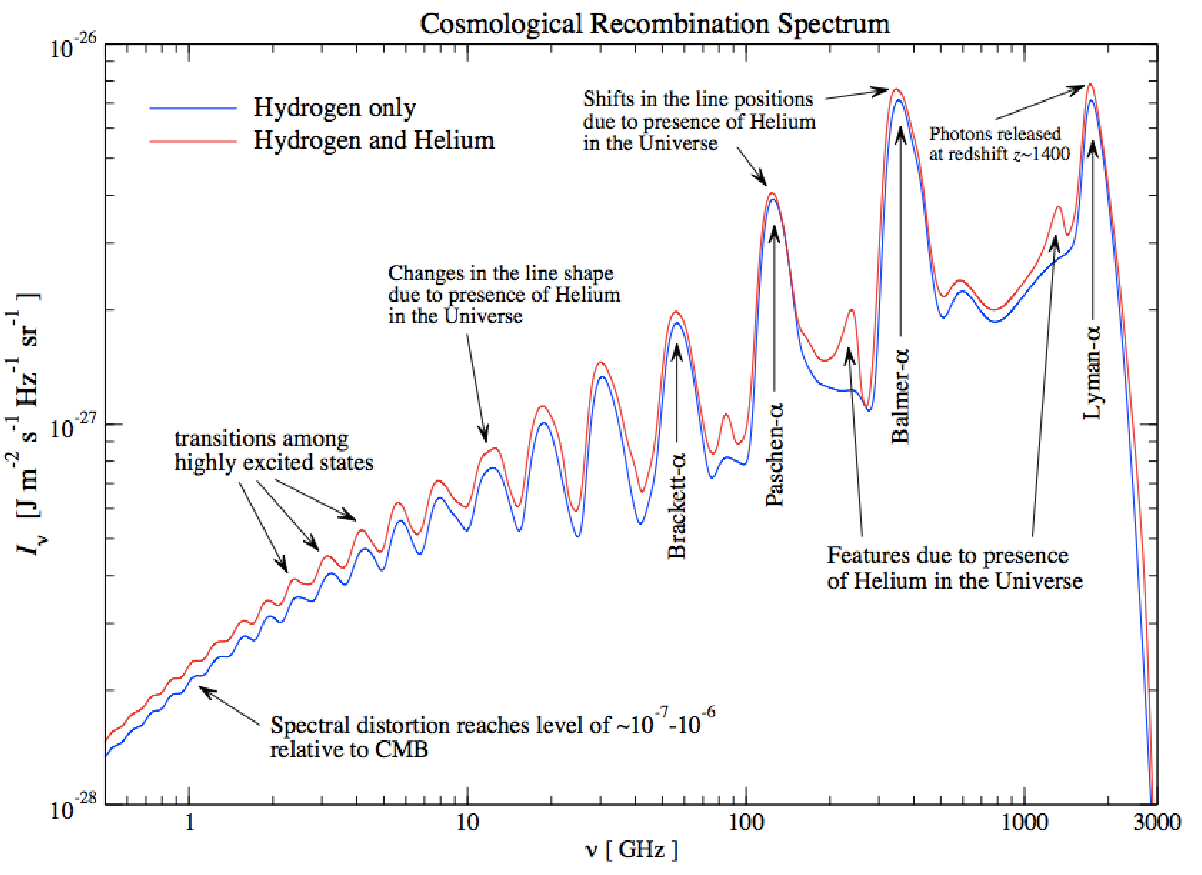}}
\caption{\label{herecfig} Intensity of  the recombination lines from
  the epoch of helium and hydrogen recombination. Figure courtesy of 
  Chluba, Rubino-Martin and Sunyaev based on calculations in  
 Ref.~\protect\refcite{cs2006b,rcs2008}.}
\end{figure}
One of the most important phase transitions in the history of the Universe
is recombination of electrons with protons and helium nuclei to form
neutral atoms \cite{zks68,peebles68}. The standing sound waves, excited by
primordial initial perturbations, in the
previously tightly coupled electron-baryon-photon plasma are frozen into
the free streaming photons \cite{sz1970c,Peebles1970} and are observed today by CMB
experiments such as COBE/DMR \cite{cobedmr}, BOOMERANG \cite{BOOMERANG},
ACBAR \cite{acbar}, WMAP \cite{wmap}, SPT \cite{spt}, ACT \cite{act} and
many others.\footnote{A more complete list of CMB experiments is available
  on NASA LAMBDA website, \url{http://lambda.gsfc.nasa.gov/product/expt}}
The recombination of HeIII to HeII at redshift $z\approx 5900$, HeII to HeI at
$z\approx 2000$ and HII to HI at $z\approx 1300$ is accompanied by the release
of recombination radiation from electrons cascading down to the ground
states of recombining atoms.  The recombination radiation from    
  Ly-$\alpha$ and $2s-1s$ 2-photon transitions   was
first mentioned  by
  Kurt, Zeldovich and Sunyaev \cite{zks68} and Peebles \cite{peebles68}
  while Dubrovich \cite{d1975} pointed out the importance of $(n,n-1)$ transitions
in the hydrogen atoms for the observations of the recombination spectrum. The full recombination spectrum was calculated
in Refs~\refcite{cs2006b,rcs2006,rcs2008}. Fig. \ref{linevisfig} shows the
line profiles for the recombination lines. It is interesting to note
that for hydrogen most of  the recombination photons are emitted
significantly 
earlier than the last scattering surface (where the Thomson visibility
function is peaked). Fig. \ref{hrecfig} shows the hydrogen recombination
spectrum and Fig. \ref{herecfig} demonstrates the contributions from helium to the
total recombination spectrum. The x-axes shows the observed frequency today
which is redshifted by a factor of $\sim 1300$ from the rest frame frequencies
and appear today in the radio part of the electromagnetic
spectrum. Lyman, Balmer and higher series lines are easily
identifiable. Approximately $\sim 5$ photons per hydrogen atom are emitted.
Presence of helium adds additional spectral features in the
recombination spectrum in Fig. \ref{herecfig}; these features would provide
a completely independent measure of the primordial helium abundance if the
cosmological recombination lines will be  detected. Proposed experiment PIXIE
\cite{pixie} would have the sensitivity to detect the CMB recombination
spectrum at $5-6\sigma$ \cite{kogut}.
 
\section{Frequency dependent blurring of the CMB anisotropies from resonant
  scattering by metals and a way to measure the abundance of metals during the
  epoch of reionization}
\begin{figure}
\resizebox{\hsize}{!}{\includegraphics{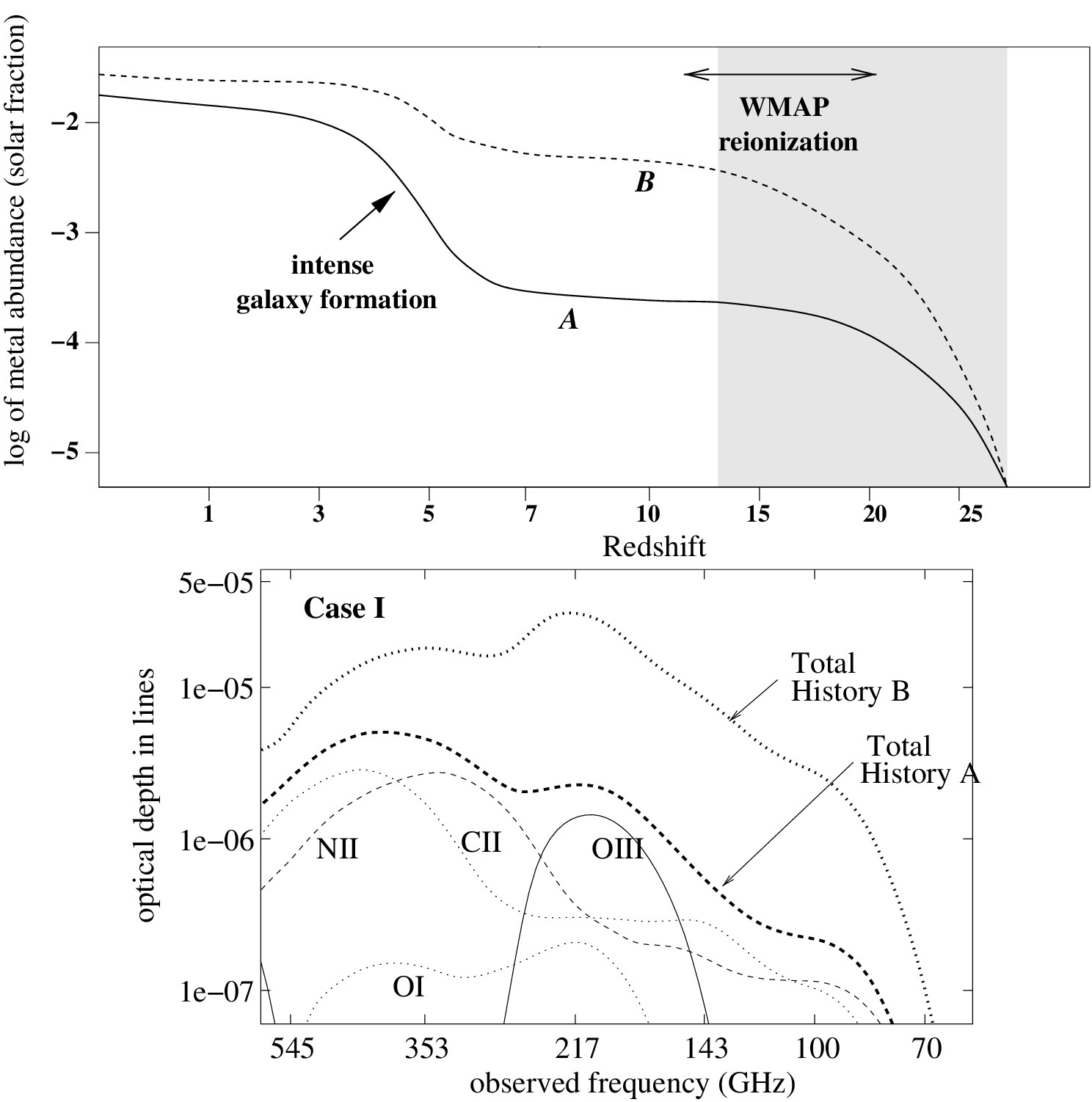}}
\caption{\label{basufig} Optical depth to the last scattering surface from
  resonant scattering on metals for the fine structure transitions
    of  NII ($205.3~\mu{\rm  m}$), OI ($63.18~\mu{\rm  m}$), OIII ($88.36~\mu{\rm
      m}$) and CII ($157.74~\mu{\rm  m}$) lines, for two different histories of ionization
  and metal production, are shown. Individual contributions are also shown for history
  A. Figure is taken from Ref~\protect\refcite{basu}.}
\end{figure}

\begin{figure}
\resizebox{\hsize}{!}{\includegraphics{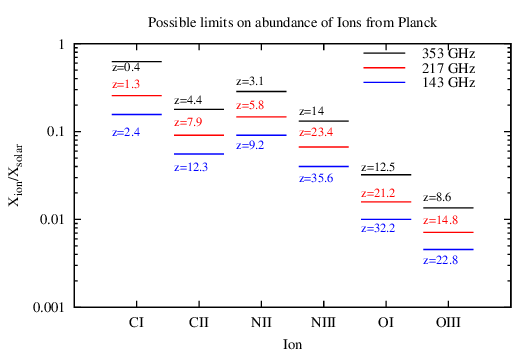}}
\caption{\label{metals} Possible limits on metals from Planck if the
  difference in brightness between $100$GHZ channel and $353,217$ and $143$GHz
  channels  of $\sim 0.01~\mu{\rm K}$ for hot/cold spots with $\Delta
  T=100~\mu{\rm K}$ ($0.01\%$ measurement) will be detected.} 
\end{figure}

Planck mission \cite{planck} will provide us with  sensitive independent measurements of
the cosmic
microwave background (CMB) angular fluctuations in different spectral
bands. Using just the existing Planck channels, we can find upper limits (and possibly a
measurement) of the abundance of several important ions, such as OIII, OI, CII, NII etc, at
different redshifts. The sensitivity to the presence of metals between the
last scattering surface (LSS) and us comes from the resonant scattering of the primordial
angular fluctuations of CMB by atoms and ions in the intergalactic space
between halos and regions inside halos with density smaller than the
critical density, when collisions are unimportant \cite{basu}. This resonant line  scattering
by atoms and ions, just like Thomson scattering from electrons, blurs the
CMB anisotropies on scales smaller than horizon,
\begin{align}
\left.\frac{\Delta T}{T}\right|_{z=0}(\nu,\mathbf{\hat{n}})\approx e^{-\tau_{\rm LSS}(\nu)}\left.\frac{\Delta T}{T}\right|_{\rm LSS}(\mathbf{\hat{n}}),
\end{align}
where $\left.\frac{\Delta T}{T}\right|_{z=0}(\nu,\mathbf{\hat{n}})$ is the observed temperature
anisotropy in CMB today in the direction $\mathbf{\hat{n}}$, $\left.\frac{\Delta
  T}{T}\right|_{\rm LSS}(\mathbf{\hat{n}})$ is the anisotropy we would see if the
optical depth to the last scattering surface $\tau_{\rm LSS}(\nu)$ was
zero, i.e. there was no reionization. Formally, this solution arises as the
boundary term in the line of sight integral solution  of the first order
Boltzmann equation of photons \cite{cmbfast,dod}. The total optical depth
has a frequency independent part arising from Thomson scattering\cite{wmap7}, $\tau_{\rm
  T}\approx 0.087$ and a frequency dependent part arising from the scattering
with metals, $\tau_X(\nu)$. $\tau_X(\nu)$ was calculated for almost all 
the important metal species in Ref.~\refcite{basu}. Figure \ref{basufig}
from Ref~\refcite{basu} shows the total optical depth as well as
the contributions from individual metal species for two different models of
ionization and metal production.  The effect of the non-zero optical depth to the
LSS is to blur the anisotropies, i.e. the hot spots become colder and the cold
spots become warmer. The angular power spectrum is  suppressed by a
factor of $e^{-2\tau}$. The best
signal-to-noise in the power spectrum is, of course, obtainable for the
anisotropies around the first acoustic peak of CMB, around angular
wavenumber $\ell \approx 220$. We can of course use all the CMB fluctuation data
below the horizon size at the redshift of scattering, $\ell\gtrsim 10-30$
and up to the angular resolution limit of Planck $\ell \lesssim 2500$, to
improve the sensitivity. Alternatively, we can stack up all the hot and
cold spots in the cleanest portions of the sky and measure the difference
in their temperatures at different frequencies, $\left<\Delta T/T_{\rm
  hot/cold}(\nu)\right>$. This analysis would be similar to the one done by
the WMAP team to detect polarization generated at the last scattering surface
\cite{wmap7}. They found $12,387$ hot spots and $12,628$ cold spots with
rms temperature fluctuation of $83.9$ $\mu$K. In our case, the resonant scattering
of the radiation will decrease the spot brightness of hot spots at the
frequency of observation, $\nu_{\rm obs}$, if there is significant amount
of the corresponding ion at redshift $(1+z)=\nu_{\rm res}/\nu_{\rm obs}$,
where $\nu_{\rm res}$  is the rest frame frequency of the resonant
line. Similarly, resonant scattering will decrease the amplitude $\Delta T$
of the cold spots, increasing their brightness. Fig. \ref{basufig} shows
that resonant scattering is unable to blur the fluctuations at low
frequencies, because they correspond to high redshifts ($z>30$) where there are
negligible amount of metals,  for the
lines of most abundant ions. By comparing the brightness of hot and cold spots
at different frequencies, we have an opportunity to measure the abundance
of ions during the epoch of reionization.  The analysis is
limited not by sensitivity but by the calibration error. A detection of an optical
depth of $10^{-4}$ using the hottest and coldest spots with temperature
fluctuation $\sim 100\mu$K would require a calibration accuracy of $\sim
0.01~\mu$K. Any difference between two frequency channels
above the calibration error would be due to resonant scattering and will constrain the presence of
metals between us and the last scattering surface. Possible limits are shown in
Fig. \ref{metals} for the most important metal species.  

The limits on the metal abundance obtained in
this way from the \emph{blurring effect} are density averaged limits.  Sensitivity of Planck in the HFI  channels
$\nu=100,143,217~{\rm GHz}$ is $\sim 2.6,0.7,1.4(\mu K)^2$ respectively at $\ell=220$ using just $50\%$
of sky for 14 months of operation \cite{planck} and much better when using
all the data and  averaging over all $\ell$ modes. Planck thus has, in principle,
opportunity to get independent upper limits to the difference in optical
depth to LSS seen by different frequency channels, if the frequency channels
can be calibrated relatively to each other precisely. The most promising
method of calibration is by using the orbital dipole of the Planck
spacecraft. The motion of the earth and Planck satellite in the solar
system can be measured with exquisite precision. In fact, for the WMAP
satellite, the orbital dipole is known at a precision of $\sim 0.1$nK
\cite{wmap3} and it is thus, in principle, possible to achieve a calibration
precision at the  level of $10^{-6}$ for the hottest/coldest spots with
temperature fluctuation of $100~\mu$K, improving the numbers presented in
Fig. \ref{metals} by a factor of 100! An additional precise source
   of calibration with well defined spectrum is also provided by the $y$-type distortion quadrupole
  induced by our motion with respect to the CMB\cite{kk2003,cs2004,ks2013}.
 
\section{$y$-type spectral distortions}
The primordial plasma of our Universe is of very low density. Average
number density of photons exceeds that of the electrons by a factor of $10^9$.
Under these circumstances, Compton scattering of radiation on free thermal
electrons (taking into account the Doppler shift of  photon energy,
the recoil effect and double Compton scattering) is the most important physical
process responsible for the interaction of matter and radiation in the
Universe. Our Universe is optically thin to bremsstrahlung absorption and
emission of CMB  photons\cite{zs1969} almost up to the time of positron-electron
annihilation which occurred at redshifts $z \sim  10^9$.
\begin{figure}
\resizebox{\hsize}{!}{\includegraphics{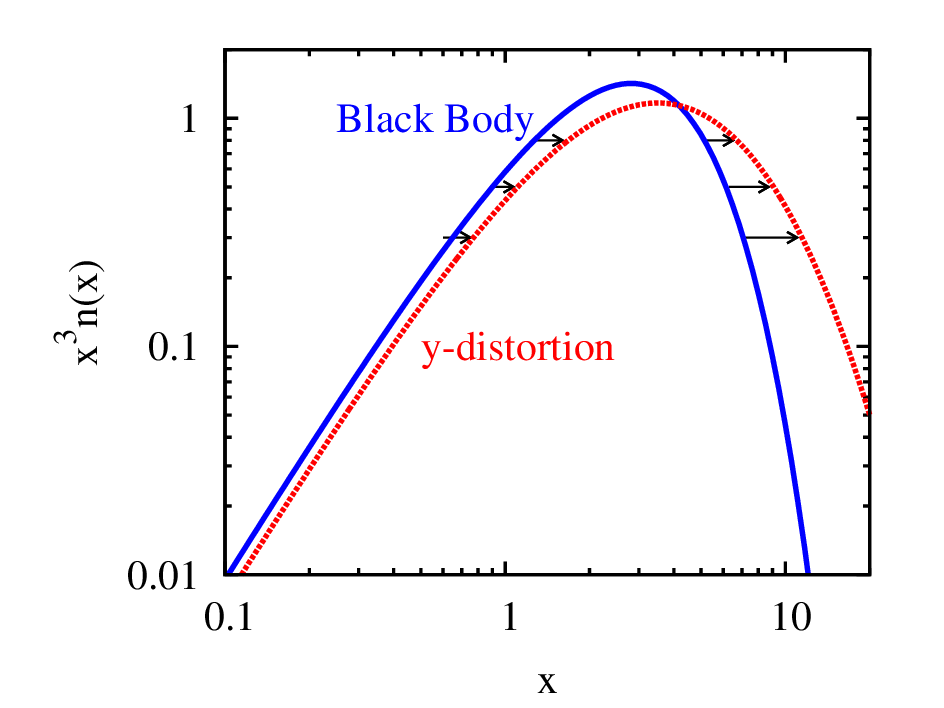}}
\caption{\label{yszfig} $y$-type distortion. Compton scattering  of CMB blackbody
  photons  with hot thermal electrons up-scatters low frequency
  photons to higher frequencies \cite{zs1969}. Occupation number $n(x)$ multiplied by
  $x^3$, making it proportional to intensity, as a function of
  dimensionless frequency $x=h\nu/\kB T$ is shown, where $T$ is the
  temperature of the blackbody spectrum. This effect\cite{sz1972} has been  observed by
  several instruments including Planck, ACT and SPT in the direction of
  many clusters of galaxies.}
\end{figure}
\begin{figure}
\resizebox{\hsize}{!}{\includegraphics{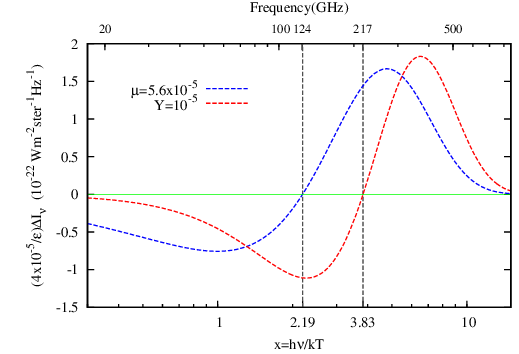}}
\caption{\label{ymufig} $y$-type and $\mu$-type distortions. Difference in
  intensity from the blackbody spectrum is shown. The distortions shown
  correspond approximately to the COBE/FIRAS limits. Both types of
  distortions shown have the same energy density and number density of
  photons. Figure from Ref.~\protect\refcite{ks2012b}.}
\end{figure}

The interaction of CMB blackbody photons (temperature $T$) with the plasma
 (temperature $\Te$) is described by the Kompaneets equation \cite{k1956}.
 The Kompaneets equation is the Fokker-Planck
approximation of Boltzmann equation with Compton scattering,
\begin{align}
\frac{\partial n}{\partial \yg}=\frac{1}{x^2}\frac{\partial }{\partial
  x}x^4\left(n+n^2 +\frac{T_e}{T}\frac{\partial n}{\partial x}\right),\label{komp}
\end{align}
where we have defined 
\begin{align}
\yg(z,z_{\rm{max}})=-\int_{z_{\rm{max}}}^{z}dz\frac{k_B\sigma_T}{m_e
  c}\frac{n_eT}{H(1+z)},\label{yz}
\end{align}
$x=h\nu/(\kB T)$ is the dimensionless frequency, $h$ is the Planck's
constant, $\nu$ is frequency, $\kB$ is the Boltzmann's constant. Also, $H(z)$ is the Hubble parameter, $\Ne$ is the electron number density and
$\zmax$ is the maximum redshift at which we start the calculation.
 The
three terms in the right-hand side brackets in Eq. \eqref{komp} describe
the change in photon frequency due to 
recoil ($\delta \nu/\nu \sim -h\nu/(\me c^2)(1-\cos\theta)$), induced recoil
and Doppler effect ($(\delta \nu/\nu)_{\rm rms}\sim 4\kB \Te /(\me c^2)$)
respectively. The (induced) recoil and Doppler effects cancel
each other if the photon spectrum is a Planck/Bose-Einstein spectrum
 with
temperature $T$ and $\Te=T$, making the right-hand side vanish.

The $y$-type distortion  is the solution of the Kompaneets equation in the
minimal comptonization limit, $\yg\ll1$. This solution, in the small
distortion limit,  is easily obtained
by approximating the occupation number on the right hand side of
Eq. \eqref{komp}  by Planck
spectrum ($\npl=1/(e^x-1)$), and using $\npl+\npl^2 =-\frac{\partial \npl}{\partial x}$.
The resulting differential equation is now trivial to integrate, giving ( 
for $y\ll 1$)\cite{zs1969}
\begin{align}
n_{\y}(x)=y\frac{xe^x}{(e^x-1)^2}\left[x\left(\frac{e^x+1}{e^x-1}\right)-4\right]\label{linsz},
\end{align}
where the amplitude of the distortion is proportional to the electron
pressure (in the limit $\Te\gg T$), integrated along the line of sight,
\begin{align}
\y&= -\int_\zmax^\zmin \frac{k_B\sigma_T}{m_e
  c}\frac{n_e(\Te-T)}{H(1+z)}\id z\label{ypar}\\
&\approx^{\Te\gg T} -\int_\zmax^\zmin \frac{k_B\sigma_T}{m_e
  c}\frac{n_e\Te}{H(1+z)}\id z\equiv y_{\rm e}\label{ye},
\end{align}
For constant temperature $\Te$, $y$ is proportional to the integrated
Thomson optical depth, $y\approx \frac{\kB \Te}{\me c^2} \tau=1.7\times
10^{-10} \Te \tau$. If the total energy injected into the CMB is $\Delta
  E/E_{\gamma}$, where $E_{\gamma}$ is the energy density of the CMB, then the
  $y$-type distortion is given by \cite{zs1969}
  $y=(1/4) \Delta
  E/E_{\gamma}$.
Eq. \eqref{ypar} also shows that we have a
positive distortion for $\Te>T$ and a negative distortion for
$\Te<T$. $y$-type distortion is shown in Figs. \ref{yszfig} and \ref{ymufig}
\subsection{$y$-type distortions from reionization and WHIM}
\begin{figure}
\resizebox{\hsize}{!}{\includegraphics{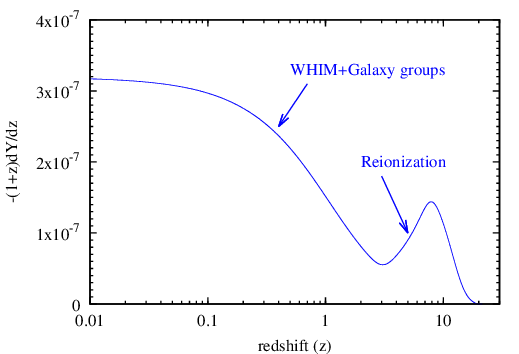}}
\caption{\label{yri} Sky averaged thermal $y$-type distortions created after
  recombination. Initially the gas is colder than the CMB because
  non-relativistic baryons cool faster as $\Te \propto (1+z)^2$ compared to
  photons $T\propto (1+z)$\protect\cite{zks68,peebles68,cs2011}. Once first stars form and
  reionization starts, gas is heated above the CMB temperature and much
  larger positive $y$-type distortions are created. $\id y/\id {\ln(1+z)}$
  is plotted which is approximately equal to the $y$-type distortions
  created in redshift interval $\delta z\sim z$. It is likely that the
  $y$-type distortions from WHIM would dominate over the reionization
  contribution. This figure is taken from Ref~\protect\refcite{ks2012b}.}
\end{figure}
Present observations indicate that the Universe was reionized between
redshifts of $6\lesssim z \lesssim 20$ \cite{wmap7}, when the first stars
and galaxies flooded the Universe with ultraviolet radiation. The ionizing
radiation also heated the gas to temperatures well above the CMB
temperature, with the electron temperature in the ionizing regions $\Te\sim
10^4~{\rm K}$. Late time structure formation shock heated the gas to even higher
temperatures \cite{sz1972b}, $10^5\lesssim \Te\lesssim 10^7~{\rm K}$,
creating the warm-hot
intergalactic medium (WHIM)\cite{co1999}.  The optical depth, $\tau$, to the last scattering surface
is well constrained by CMB observations \cite{wmap7} to be $\tau \approx
0.087\pm 0.014$, assuming $\Lambda$CDM cosmology. Thus if $\Te\approx
10^4~{\rm K}$, we expect $y\sim 10^{-7}$. However if a significant fraction
of baryons end up in the WHIM at $z\lesssim 3$, as expected from recent
simulations \cite{co1999,co2006}, we expect the $y$-distortions from the WHIM
to dominate over those from reionization \cite{ns2001}. In any case, these distortions
would be easily detected by PIXIE \cite{pixie} and the next generation CMB experiments like CoRE
\cite{core}, ACTPol \cite{actpol} and SPTPol \cite{sptpol} should also be
able to detect the fluctuations in the $y$-type distortions from the
the WHIM. The rate
of $y$-type distortion injection with redshift is shown in Fig. \ref{yri} for a simple model
where reionization happens between $8<z<15$ and the density averaged
temperature of 
free electrons is assumed to be $\Te=10^4~{\rm K}$ for $z>3$ and
$\Te=10^6/(1+z)^{3.3}~{\rm K}$ at $z<3$ \cite{co1999}. The contribution
from the WHIM to the $y$-type distortions dominates over those from
reionization in this model.

\subsection{$y$-type distortion from averaging of blackbodies in our CMB sky}
\begin{figure}
\resizebox{\hsize}{!}{\includegraphics{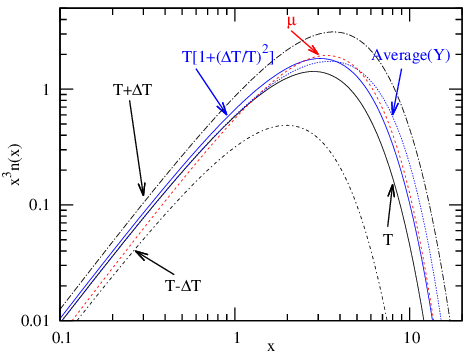}}
\caption{\label{bbmixnx} Spectrum resulting from averaging two blackbodies
  with temperatures $T\pm \Delta T$. We have used large value of $\Delta T$
to make changes visible, but have used formula with only lowest order
non-vanishing correction to the blackbody spectrum valid for small
distortions. We have plotted intensity (up-to a numerical constant),
$n(x)x^3\propto I_{\nu}$. Figure taken from Ref.\protect\refcite{ksc2012}.}
\end{figure}
\begin{figure}
\resizebox{\hsize}{!}{\includegraphics{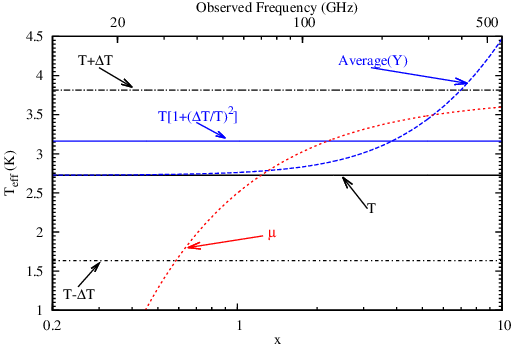}}
\caption{\label{bbmixt} Same as Fig. \ref{bbmixnx}, but effective blackbody
  temperature defined by $n(\nu,\Teff)\equiv\frac{1}{e^{h\nu/(\kB
      \Teff)}-1}$ is plotted. Figure taken from
  Ref.\protect\refcite{ksc2012}. At high redshifts $\yg$ becomes
  significant  (see Fig.
  \ref{yfig}, Eq. \ref{yz}) and comptonization converts $y$-type distortions created by
  the mixing of blackbodies to intermediate-type (or $i$-type) and $\mu$-type distortions.}
\end{figure}
The temperature of the CMB in the sky is not constant but is a function of
direction. The most significant anisotropy comes from our peculiar motion
with respect to the CMB rest frame (i.e the frame in which the CMB dipole
is zero). Our peculiar velocity, or equivalently the CMB dipole seen in our rest
frame, was measured very precisely by COBE \cite{cobe} and WMAP
\cite{wmapdipole} experiments and has an amplitude $\Delta T_{\rm dipole}3.355\pm 0.008~{\rm
  mK}, v/c\approx 1.23\times 10^{-3}$. In addition, there are small scale
anisotropies of amplitude $10~\mu{\rm K}$ in the microwave sky due to the
presence of primordial density perturbations. If we
average the CMB intensity over all angles or part of the sky, either
explicitly to improve the sensitivity or because of the finite beam of the
telescope, we will inevitably mix the blackbodies of different
temperatures. It was shown in Ref.~\refcite{zis1972} that mixing of
blackbodies gives rise, at the lowest non vanishing order, to a $y$-type
distortion and that the thermal $y$-type distortion from comptonization is
just superposition of blackbodies. This result is quite general and is also
applicable if the source of electron motion is not thermal but kinetic,
i.e. there is a $y$-type distortion arising from peculiar motion of baryons. As a simple example, superposition of two
blackbodies with temperatures $T\pm \Delta T$ is shown in
Fig. \ref{bbmixnx} (intensity) and Fig. \ref{bbmixt} (effective
temperature). The resulting spectrum after averaging can be recognized as a
blackbody with higher temperature $T\left(1+\left(\frac{\Delta
      T}{T}\right)^2\right)$ with a $y$-type distortion of amplitude $ \left(\frac{\Delta
      T}{T}\right)^2/2$ \cite{cks2012,ksc2012}. For the dipole anisotropy in
  our sky, we get $y\approx 2.5\times 10^{-7}$ and averaging of the rest of
  the smaller scale
  anisotropies (without dipole) contribute $y\approx 8\times 10^{-10}$. The $y$-type
  spectrum  provides, in principle, a high precision source to
  calibrate CMB experiments \cite{cs2004}.

\section{$\mu$-type distortions}
\begin{figure}
\resizebox{\hsize}{!}{\includegraphics{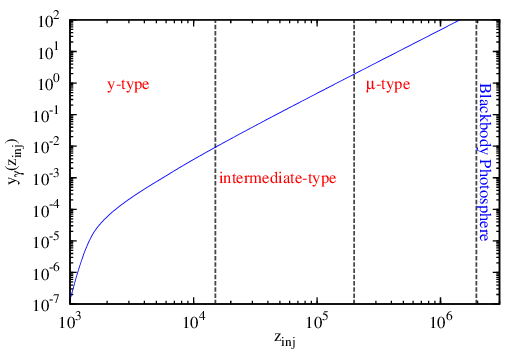}}
\caption{\label{yfig} The Compton parameter $\yg$, Eq. \ref{yz}, is plotted. At $z\gtrsim
  10^5$ $\yg>1$ creating a Bose-Einstein spectrum, unlike the amplitude of distortion $y$ which is always
  smaller than unity (as verified by COBE/FIRAS\protect\cite{cobe}). This figure is
taken from Ref~\protect\refcite{ks2012b}.}
\end{figure}

The thermal capacity of electrons and baryons is
negligible in comparison with that of photons and Compton interaction
rapidly establishes a Maxwellian distribution of electrons with temperature
$\Te$ defined by the
radiation (photon) field \cite{zl1970,ls1971}. If the  spectrum of photons
is  Bose-Einstein spectrum   then the
equilibrium electron temperature is exactly equal
to the radiation temperature. The other way around, the timescale for the
establishment of equilibrium 
Bose-Einstein  spectrum for photons through Compton scattering is
comparatively much longer compared to the age of the Universe at redshifts
$z\lesssim 2\times 10^5$, see Fig. \ref{yfig} and the discussion below.

Bose-Einstein spectrum with chemical potential parameter $\mu$\footnote{Our definition of $\mu$
  is equivalent to the negative of 
     the usual statistical physics chemical potential divided by
  temperature, making it dimensionless and we will call it \emph{chemical
    potential parameter.}} and temperature $\Te$, $n_{\rm BE}=1/(e^{h\nu/(\kB\Te)+\mu}-1)=1/(e^{xT/\Te+\mu}-1)$ is the equilibrium
solution of the Kompaneets equation, as can be readily
verified by substituting it in Eq. \eqref{komp}. The reference temperature
$T$ now corresponds to a blackbody spectrum with the same number density of
photons as the Bose-Einstein spectrum, and the factor $T/\Te$ is there
because we have defined the dimensionless frequency $x$ with reference to $T$. For small distortions,
$\mu\ll1 $, we can expand the Bose-Einstein spectrum around the reference
blackbody giving, for the $\mu$-type distortion (using\cite{is1975b} $\Te/T-1=\mu/2.19$), 
\begin{align}
n_{\mu}=\frac{\mu e^x}{\left(e^x-1\right)^2}\left(\frac{x}{2.19}-1\right).
\end{align}
The intensities of $y$ and $\mu$-type distortions are plotted in
Fig. \ref{ymufig}. The $\mu$ parameter is related to the fractional energy injected
into the CMB by\cite{sz1970,is1975b} $\mu=1.4\Delta E/E_{\gamma}$.

The $\mu$-type distortion is thus the solution of the Kompaneets equation in the
saturated comptonization limit, for $\yg\gg 1$.
The function $\yg(0,\zinj)$ is plotted in Fig. \ref{yfig}. If $\yg\gtrsim 1$,
comptonization is very efficient in establishing a Bose-Einstein spectrum
 if energy is injected at redshift $\zinj$, while for $\yg\ll 1$ only a
 $y$-type distortion can be created from the heating or cooling of the CMB.
We should note that, on the contrary, the $y$-parameter in Eq. \eqref{ypar}
describes the amplitude of injected energy/$y$-type distortion  and is
always much smaller than unity for the physical processes
  considered in this paper. 

\subsection{Negative $y$ and $\mu$ distortions from Bose-Einstein condensation of CMB}\label{becsec}
\begin{figure}
\resizebox{\hsize}{!}{\includegraphics{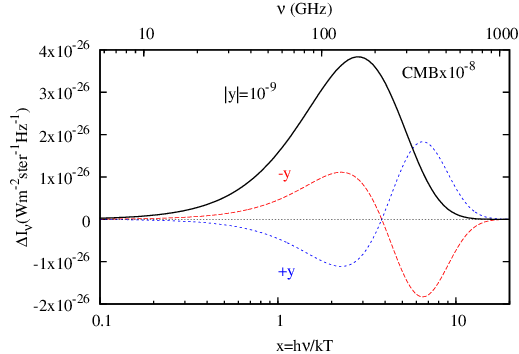}}
\caption{\label{becfig1} The negative $y$-type distortions resulting from
  cooling of CMB. The expected magnitudes of $y$ and $\mu$ parameters for
  CMB is shown in red in Tables \ref{tbl1} and \ref{tbl2}. Figure from Ref.~\protect\refcite{ksc2012}.}
\end{figure}
\begin{figure}
\resizebox{\hsize}{!}{\includegraphics{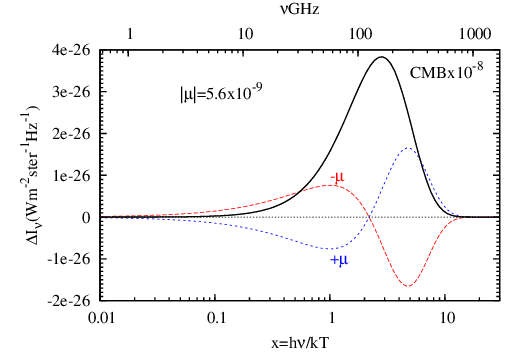}}
\caption{\label{becfig} The negative $\mu$-type distortions resulting from
  cooling of CMB. The expected magnitudes of $y$ and $\mu$ parameters for
  CMB is shown in red in Tables \ref{tbl1} and \ref{tbl2}. Figure from Ref.~\protect\refcite{ksc2012}.}
\end{figure}
It was recognized long ago \cite{zks68} that in the early Universe, when
baryons and photons are tightly coupled, the photons must transfer energy
to the
baryons to keep them in equilibrium (until $z\sim 500$), as baryons cool adiabatically faster ($\Te\propto
(1+z)^2$) than radiation ($T\propto (1+z)$) with the expansion of the
Universe. This is simply because the baryons are non-relativistic with an
adiabatic index of $5/3$ compared to the adiabatic index of $4/3$ for the
relativistic photon gas. This (small) cooling of the CMB gives rise to spectral
distortions ($y,\mu$ and intermediate-type) which are exactly the negative
of the distortions caused by the heating of the CMB. Figure \ref{becfig1} shows
the $y$-type distortion resulting from the cooling of the CMB. The $\mu$-type distortion
resulting from the cooling of the CMB is shown in Fig. \ref{becfig}  along with
a positive distortion resulting from an equivalent amount of heating.  There is, however, one very
important physical difference between the heating and the cooling of the CMB. The
cooling of the CMB results in an excess in the number density of photons compared to the
blackbody radiation with the same energy density, and this can be
recognized as the condition for the Bose-Einstein condensation of photons to
happen \cite{is1975b,ksc2012}. In the case of the CMB, of course, the
  condensing photons, moving to lower frequencies due to the stimulated
  scattering and the recoil effect,   are
efficiently destroyed by bremsstrahlung and double Compton absorption \cite{cs2011,ksc2012},
and no actual photon condensate\cite{ksvw2010} is formed.

\section{Beyond $\mu$ and $y$: Intermediate-type distortions}
\begin{figure}
\resizebox{\hsize}{!}{\includegraphics{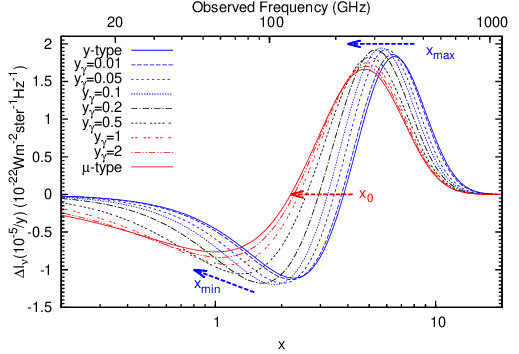}}
\caption{\label{interfig} Intermediate-type  distortions are shown for
  different values of Compton parameter $(\yg,z_{inj})=(0.01,1.56\times
  10^4), (0.05,3.33\times 10^4), (0.1,4.67\times 10^4), (0.2,6.55\times 10^4),
  (0.5,1.03\times
  10^5), (1,1.45\times 10^5), (2,2.04\times 10^5)$. Figure is taken from Ref.~\protect\refcite{ks2012b}.}
\end{figure}
The $y$-type distortions are created at redshifts $z\lesssim 1.5\times
  10^4, \yg\lesssim 0.01$, when the comptonization is minimal. At $z\gtrsim 2\times
  10^5$, the Compton parameter $\yg\gtrsim 2$ and we get $\mu$-type
  distortion, which is the solution of the Kompaneets equation, Eq. \eqref{komp} in the
  saturated comptonization limit.
The energy injected between the redshifts $1.5\times 10^4\lesssim z \lesssim
2\times 10^5$ is only partially comptonized as the Compton parameter is
$0.01\lesssim \yg\lesssim 2$, see Fig \ref{yfig}. The spectrum created from
the energy injection in
this redshift range is therefore in between $y$-type distortion (minimal
comptonization) and a $\mu$-type distortion (saturated
comptonization). Figure \ref{interfig} shows the spectrum that would be
obtained if the energy is injected at redshifts corresponding to different
$\yg$ parameters. In practice, the energy is more likely to be injected
continuously over a redshift range rather than instantaneously at a single
redshift and we expect the observed spectrum to be a linear superposition
(with appropriate weights)
of the distortions for different values of $\yg$. Since the
intermediate-type spectrum depends on the redshift of energy injection,
there is additional information here compared to the $y$ and $\mu$-type distortions
which only remember the total energy injected and not the exact time of
energy injection. The rather simple behavior of the
  intermediate-type distortions with respect to the energy injection redshift
  opens up the possibility to measure the redshift dependence of the energy
  injection rate. For example, in case of Silk damping or dark matter
  annihilation we  the energy injection rate is a power law in
  redshift, $\id Q/\id z \propto (1+z)^{\alpha}$ and with the intermediate-type
  distortion we can measure the parameter $\alpha$ and thus the spectral
  index of the primordial power spectrum. Combining the intermediate-type
  distortions and the $\mu$-type distortions also allows to distinguish between
  a power law energy injection and the exponential dependence of energy
  injection rate on the redshift expected from particle decay. Intermediate-type
distortions are explored in detail in Ref.~\refcite{ks2012b}.

\section{Blackbody photosphere of the Universe}\label{bb}
\begin{figure}
\resizebox{\hsize}{!}{\includegraphics{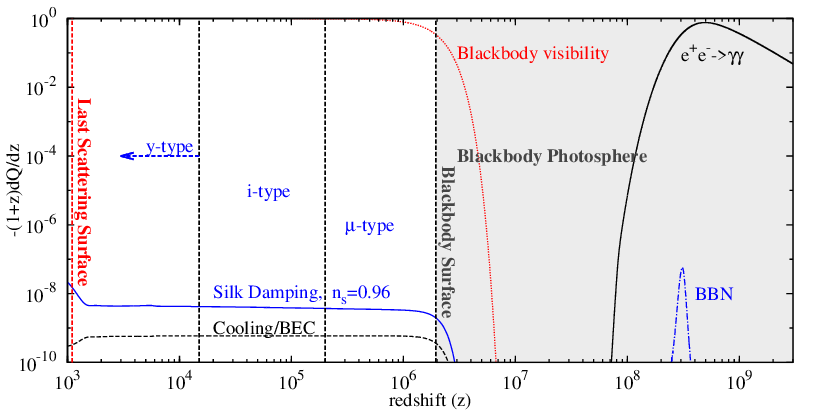}}
\caption{\label{energyfig} Comparison of Silk damping (heating)  and
  Bose-Einstein condensation (cooling) rates in standard cosmology. Also
  shown is the blackbody visibility function defining the blackbody
  photosphere, where the distortions from the blackbody are exponentially
  suppressed (see section \ref{bb} for details). Since heating of photons
  from electron-positron
  annihilation happens deep inside the blackbody photosphere, the photon
  spectrum   hardly deviates from a blackbody even though the comoving energy
  density of photons is more than doubled. This situation is to be
  contrasted with neutrinos, heating of which due to electron positron
  annihilation is just $1\%$ but it results is significant deviations from
  a Fermi-Dirac distribution which are important for big bang
  nucleosynthesis (BBN) \cite{dt1992,fdt1993,dolgov,mm2005}. Also shown is the
  nuclear binding energy released during helium production in BBN
  calculated
 using Kawano's modification\cite{kawano} of Wagner's
  code\protect{\cite{wagoner}}.}
\end{figure}
The blackbody spectrum of CMB is created dynamically in the early
Universe, initially because of the complete thermal equilibrium between the
photons
and the electrons/positrons (and other particles at even higher redshifts)
through pair creation and annihilation and bremsstrahlung. Subsequent adiabatic expansion of
the Universe preserves the blackbody spectrum, except small BEC effects
described in section \ref{becsec} above. Once $e^{\pm}$ pair production
becomes inefficient at $z\sim 10^8-10^9$, double 
Compton scattering ($\gamma + e^{-}\leftrightarrow \gamma + \gamma + e^{-}$)
and bremsstrahlung ($Z + e^{-}\leftrightarrow \gamma + Z + e^{-}$) become the dominant mechanism of
photon absorption and emission at low frequencies while Compton scattering ($\gamma + e^{-}\leftrightarrow \gamma + e^{-}$)
efficiently redistributes the photons in energy (comptonization). The redistribution of energy
among the available photons by comptonization establishes a Bose-Einstein
spectrum with a chemical potential parameter $\mu$ at $z\gtrsim 10^5$ while
the emission/absorption of photons by double Compton and bremsstrahlung drives
the chemical potential parameter to zero at $z\gtrsim 10^6$ creating a Planck
spectrum. 

Understanding the creation of the blackbody spectrum therefore requires
solving the Kompaneets equation \cite{k1956}, describing comptonization, with
the source terms arising from bremsstrahlung and double Compton scattering
\cite{lightman,thorne81}.
 A quite accurate solution to this  partial differential
equation for comptonization with a source term responsible for the emission and
absorption of photons was  found analytically in Ref.~\refcite{sz1970}. Although
Ref.~\refcite{sz1970} only considered bremsstrahlung as the
emission/absorption mechanism, the double Compton
scattering cross section is similar enough to bremsstrahlung that their
method of solution could be applied immediately to the double Compton
\begin{figure}
\includegraphics[width=9cm]{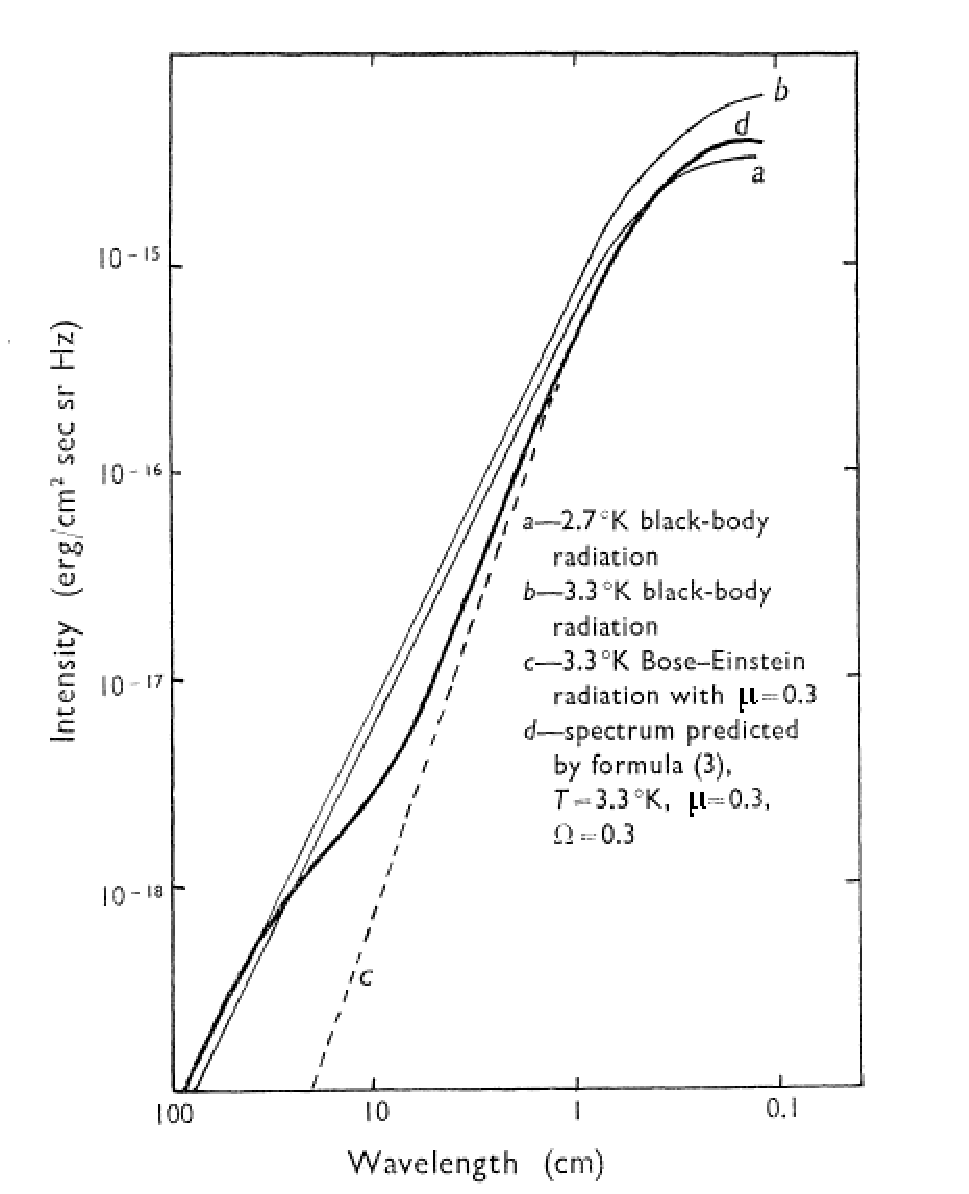}
\caption{\label{sz1970fig}This is the ancient way to describe the shape of the $\mu$-type distortions including the low  frequency part where
bremsstrahlung and double Compton effects create Rayleigh-Jeans spectrum
with temperature equal to the electron temperature 
defined by the interaction with $\mu$-type radiation field\protect{\cite{sz1970d}}.}
\end{figure}
scattering also. This was done in Ref.~\refcite{dd1982} for a low baryon
density Universe such as ours, where double Compton scattering dominates
over bremsstrahlung. Recently, 
corrections to this solution were computed in Ref.~\refcite{ks2012} and a
solution with an accuracy of $\sim 1\%$ including both bremsstrahlung and
double Compton processes was presented. The solution describes the
evolution of the chemical potential parameter of the Bose-Einstein spectrum created
at redshift $z$ and is given by
\begin{align}
\mu(z=0)&=\mu(z)e^{-\mT}\nonumber\\
\mT(z)&\approx
\left[\left(\frac{1+z}{1+\zdC}\right)^{5/2}\right]\label{dceq}
\end{align}
where, 
$\zdC \approx 1.96\times 10^6$ defines \emph{the blackbody surface}, behind which
$\mu$ is exponentially suppressed. The Bose-Einstein spectrum with
  the effect of this exponential suppression of chemical potential at low
  frequencies is shown in Fig. \ref{sz1970fig}.
The precise analytic solution for the blackbody visibility function derived
in Ref. \refcite{ks2012} $\mG=e^{-\mT}$, including the effects of
bremsstrahlung and double Compton, is
plotted in Fig. \ref{energyfig}, curve 'd'. Thus any perturbations away from the Planck
spectrum are suppressed exponentially at $z\gtrsim 2\times
10^6$. The electron-positron annihilation at $z\sim 10^8-10^9$ more than
doubles the entropy and energy   in photons, but the deviations resulting from the Planck spectrum
never rise above a tiny value of $\sim 10^{-178}$ \cite{ks2012}. This
demonstrates how difficult, almost impossible, it is to create deviation
from 
the Planck spectrum
at $z\gtrsim 2\times 10^6$, in the blackbody photosphere. 

\section{Energy release in the early Universe}

\subsection{Spectral distortions from Silk damping: A view of inflation
  spanning 17 e-folds!}
Primordial adiabatic perturbations excite standing sound waves on entering
the horizon in the early
Universe in the tightly coupled electron-baryon-photon fluid \cite{lifshitz,sz1970c,Peebles1970}.
\begin{figure}
\includegraphics[width=11cm]{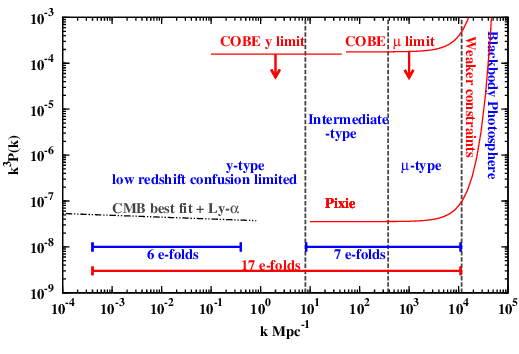}
\caption{\label{powerfig}Spectral distortions can deliver $7$ additional
  e-folds extending our view of inflation from $6$ (at present) to $17$
  e-folds. Constraints on the primordial power spectrum of initial
  curvature perturbation $P(k)$ from CMB \cite{wmap,spt,act} and
  Ly-$\alpha$ observations of SDSS \cite{ly2006,ssm2006} are shown along
  with current constraints from COBE spectral distortion measurements \cite{cobe} and
  future constraints from the proposed experiment Pixie\protect{\cite{pixie}}.}
\end{figure}
On small scales, there is shear viscosity and thermal conductivity in the
fluid which
dissipates the energy in the sound waves, suppressing the fluctuations
\cite{silk,Peebles1970,kaiser}. Microscopically, photon diffusion through
the plasma creates a local quadrupole, which is dissipated by shear
viscosity, and relative motion between the photons and the baryons creates a
local dipole, which is dissipated by thermal conduction or Compton
drag. This Silk damping of primordial fluctuations is already observed by
  the current 
\begin{figure}
\includegraphics[width=11cm]{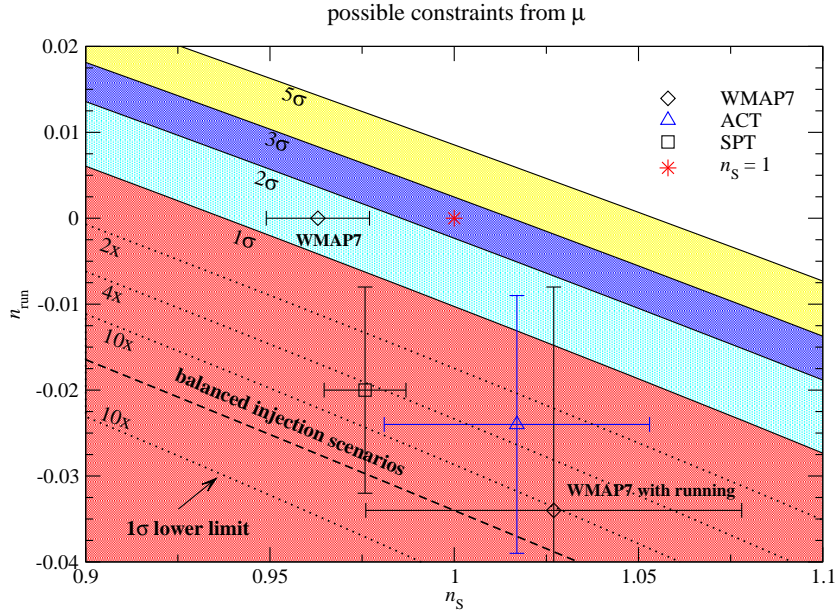}
\caption{\label{muconsfig}Possible constrains with PIXIE on spectral index
  and its running. $1-5\sigma$ limits are shown along with the $1-\sigma$
  limits if the sensitivity of PIXIE is improved by a factor of
  $2-10$. Best fit parameters from WMAP \cite{wmap7}, ACT \cite{act} and
  SPT \cite{spt} are also shown. Figure is taken from  Ref.~\protect\refcite{cks2012}.}
\end{figure}
CMB experiments SPT\cite{spt} and ACT\cite{act} with high precision. The
power which disappears from the fluctuations appears in the average CMB
spectrum or monopole as $y$, $\mu$ and intermediate-type spectral
distortion
\cite{sz1970b,daly1991,hss94,cs2011,ksc2012,cks2012,ksc2012b,pz2012}. In 
Fig. \ref{energyfig}, 
The heating rate of CMB from sound wave dissipation is compared with the cooling
due to the  energy transfer to baryons for several different power spectra
allowed by the current CMB data \cite{wmap7}. It is interesting to note that
the cases where cooling dominates over heating for $\mu$-type distortions are
still allowed if there is non-zero running of the spectral index, although
for most of the region of the parameter space heating
dominates over cooling. These distortions can be added together linearly
since they are very small. The $\mu$-type and the intermediate-type spectral
distortions are created by the dissipation of modes with comoving wavenumbers
$8\lesssim k\lesssim 10^4~{\rm Mpc}^{-1}$.  Most of this range of  scales
is inaccessible to any other cosmological probe. When combined with the
  information from the CMB anisotropies and the large scale structure, this gives
  us a  view of inflation spanning $17$ e-folds compared with $6-7$
  e-folds ($4\times 10^{-3}~\rm{Mpc}^{-1}\lesssim k \lesssim
  0.2~\rm{Mpc}^{-1}$) without the spectral distortions, as depicted in
  Fig. \ref{powerfig}.
The CMB spectral distortions
are thus measuring the primordial power spectrum on extremely small 
 scales and are  very important to our understanding of the
physics of the initial conditions. Possible constraints
from PIXIE are shown in Fig. \ref{muconsfig}.
Recently, precise calculations of these distortions were done in
Ref.~\refcite{cks2012} including previously ignored effects such as second
order Doppler effect and fitting formulae for $y$ and $\mu$-type parameters
were also provided. Distortions for several other non-standard and general
initial conditions were calculated in Refs.~\refcite{dent2012,ceb2012} and
the effect of non-Gaussian initial conditions on $\mu-T$ correlations was
pointed out in Refs.~\refcite{pajer2012,ganc2012}.

\subsection{A census of unavoidable $\mu$ and $\y$-type
distortions in standard cosmology}

A census of unavoidable $\mu$ and $\y$-type
distortions in standard cosmology is given in Tables \ref{tbl1} and
\ref{tbl2}. Public codes KYPRIX\cite{pb2009} and
CosmoTherm\cite{cs2011}  are now available to calculate the evolution
of CMB spectral distortion by numerically solving the Kompaneets
  equation with 
  bremsstrahlung and double Compton terms, starting in the blackbody photosphere at
$z\gtrsim 2\times 10^6$.  Many of the processes shown in Tables \ref{tbl1}
and \ref{tbl2} were first calculated in Ref.~\refcite{cs2011}. The CosmoTherm code also includes high precision
calculations of Silk damping and other important cases of energy injection.
A fast and precise  Mathematica code to calculate the spectral
  distortions for user-defined energy injection rates, taking advantage of  the analytic
solution described above and
pre-computed numerical templates described in the next section, is also publicly available\footnote{\url{http://www.mpa-garching.mpg.de/~khatri/idistort.html}}.

We stress here again the importance of blackbody photosphere, see
Fig. \ref{energyfig}.   The  CMB spectrum does not carry  any information
about  even strong energy
injection deep inside the blackbody photosphere. Any injected energy behind
the blackbody surface, $z\gtrsim 2\times 10^6$, is
almost completely  thermalized without any observable traces in the CMB
spectrum. The best example of the strength of equilibrium restoring processes
 (Compton and double Compton scattering,  bremsstrahlung) at high redshifts
 is the huge energy release, $\Delta E/E \sim \mathcal{O}(1)$, from electron positron
 annihilation. The    resulting deviations from the complete equilibrium
 (blackbody, $\Te=T_{\gamma}=T_{\rm ions}$)
 even in this extreme case 
 are just $10^{-178}$!.
\begin{table}
\tbl{Census of  $\mu$ distortions  in
  standard cosmology.  The adiabatic cooling of
matter results in negative distortions shown in red. Table is taken from
Ref.~\protect\refcite{ks2012}.} 
{\begin{tabular}{|c|c|}
\hline
 Process  & $ \mu$ \\
\hline
\color{blue} electron-positron annihilation & \color{blue}$  10^{-178}$\color{black}\\ 
\color{blue} BBN tritium decay & \color{blue}$ 2\times  10^{-15}$\color{black}\\ 
\color{blue} BBN $^7{\rm Be}$ decay & \color{blue}$ 10^{-16}$\color{black}\\ 
\color{blue} WIMP dark matter annihilation & \color{blue}$ 3\times
10^{-9}f_{\gamma}\frac{10{\rm GeV}}{\mdm}$\color{black}\\ 
\color{blue} Silk damping & \color{blue}$ 10^{-8} - 10^{-9}$\color{black}\\ 
\color{red}Adiabatic cooling of matter and & \color{black}\\ 
\color{red}Bose-Einstein condensation & \color{red}$ -2.7\times 10^{-9}$\color{black}\\ 
\hline
\end{tabular}\label{tbl1}}
\end{table}

\begin{table}
\tbl{Census of $y$-type distortions  in
  standard cosmology.  $y$-type distortion
  from the mixing of blackbodies in our
  CMB sky \cite{cs2004} are also shown. Adiabatic cooling of
matter creates negative distortions shown in red. Reionization/WHIM
contributions after recombination dominate. Table is taken from Ref.~\protect\refcite{ks2012}.} 
{\begin{tabular}{|c|c|}
\hline
 Process  & $y$ \\
\hline
\color{blue} WIMP dark matter annihilation & \color{blue}$ 6\times
10^{-10}f_{\gamma}\frac{10{\rm GeV}}{\mdm}$\color{black}\\ 
\color{blue} Silk damping & \color{blue}$ 10^{-8} - 10^{-9}$\color{black}\\ 
\color{red}Adiabatic cooling of matter and & \color{black}\\ 
\color{red}Bose-Einstein condensation & \color{red}$ -6\times 10^{-10}$\color{black}\\ 
\color{blue}Reionization/WHIM & \color{blue}$  10^{-6}-10^{-7}$\color{black}\\ 
\color{blue}Mixing of blackbodies: CMB $\ell\ge 2$ multipoles & \color{blue}$  8\times 10^{-10}$\color{black}\\ 
\hline
\end{tabular}\label{tbl2}}
\end{table}
\subsection{Constraining  new fundamental physics with spectral distortions}
Most of the talk  has been concentrated on the spectral distortions expected
in standard cosmology. The physics  beyond the standard model provides
numerous new possible sources of heating for CMB. In particular, it is clear from
Tables \ref{tbl1} and \ref{tbl2} that in standard cosmology we expect $\mu$
and intermediate-type distortions with an amplitude of $10^{-8}-10^{-9}$.  A
distortion of larger magnitude, if detected, would thus undoubtedly be a signal
of new physics. In addition, the most important standard source of energy
injection, Silk damping, has a power law dependence on redshift, resulting
in almost an equal amount of energy in $\mu$-type and intermediate-type
distortions \cite{ks2012b}. Thus if one of the $\mu$-type or the intermediate
type distortion is stronger than the other, it would again signify new
physics. Some of the  possible sources of  energy injection from new physics are
dark matter decay and annihilation \cite{fengreview,fengsusy,fengkk}, decay
of cosmic strings and other topological defects,  cosmic string wakes and
oscillating superconducting cosmic strings
\cite{csbook,vilenkin1988,tsv2012}, photon-axion
inter-conversion \cite{mrs2005,chrr2011}, violation of reciprocity relation
\cite{ellis2013}, dissipation of primordial magnetic
fields \cite{jko2000}, quantum wave function collapse
\cite{lochan2012},
and evaporating primordial black holes
\cite{tn04}. 

\section{The future looks good for CMB spectroscopy}
Three proposed next generation CMB experiments PIXIE \cite{pixie},
COrE \cite{core} and LiteBIRD \cite{litebird} would have the sensitivity to detect most of the spectral
features considered in this review. Current specifications of PIXIE, in
particular, with 400 frequency channels and absolute calibration, are
optimal to detect the global $y$, $\mu$ and intermediate-type distortions. COrE
and LiteBIRD would have the sensitivity and the angular resolution  to detect metals
during the epoch of reionization and angular fluctuations in the $y$-type
sky from the warm-hot intergalactic medium. In the near future, therefore, we
expect many interesting results from the observations of the spectral features
in the CMB.

\bibliographystyle{ws-ijmpd}
\bibliography{sunyaev_plenary}

\begin{thebibliography}{10}

\bibitem{cobe}
D.~J. {Fixsen}, E.~S. {Cheng}, J.~M. {Gales}, J.~C. {Mather}, R.~A. {Shafer}
  and E.~L. {Wright}, {\em \apj} {\bf 473}  (1996)   576.

\bibitem{fm2002}
D.~J. {Fixsen} and J.~C. {Mather}, {\em \apj} {\bf 581}  (2002) 817.

\bibitem{pixie}
A.~{Kogut}, D.~J. {Fixsen}, D.~T. {Chuss}, J.~{Dotson}, E.~{Dwek},
  M.~{Halpern}, G.~F. {Hinshaw}, S.~M. {Meyer}, S.~H. {Moseley}, M.~D.
  {Seiffert}, D.~N. {Spergel} and E.~J. {Wollack}, {\em \jcap} {\bf 7}  (2011)
  ~25.

\bibitem{core}
{The COrE Collaboration}, {\em ArXiv e-prints}   (2011)
  \href{http://arxiv.org/abs/1102.2181}{{\ttfamily arXiv:1102.2181
  [astro-ph.CO]}}.

\bibitem{litebird}
{M. Hazumi et al.}, { {LiteBIRD: a small satellite for the study of B-mode
  polarization and inflation from cosmic background radiation detection}} {\em
  Society of Photo-Optical Instrumentation Engineers (SPIE) Conference Series}
  {\bf 8442} (September 2012), p. 844219.

\bibitem{sz1970c}
R.~A. {Sunyaev} and Y.~B. {Zeldovich}, {\em \apss} {\bf 7}  (1970) 3.

\bibitem{jw1985}
B.~J.~T. {Jones} and R.~F.~G. {Wyse}, {\em \aap} {\bf 149}  (1985) 144.

\bibitem{wmap1}
{G. Hinshaw et al.}, {\em \apjs} {\bf 148}  (2003) 135.

\bibitem{sc2009}
R.~A. {Sunyaev} and J.~{Chluba}, {\em Astronomische Nachrichten} {\bf 330}
  (2009)   657.

\bibitem{cs2006b}
J.~{Chluba} and R.~A. {Sunyaev}, {\em \aap} {\bf 458}  (2006) L29.

\bibitem{rcs2008}
J.~A. {Rubi{\~n}o-Mart{\'{\i}}n}, J.~{Chluba} and R.~A. {Sunyaev}, {\em \aap}
  {\bf 485}  (2008) 377.

\bibitem{zks68}
Y.~B. {Zeldovich}, V.~G. {Kurt} and R.~A. {Sunyaev}, {\em Zh. Eksp. Teor. Fiz.}
  {\bf 55}  (1968)   278.

\bibitem{peebles68}
P.~J.~E. {Peebles}, {\em \apj} {\bf 153}  (1968)  ~1.

\bibitem{Peebles1970}
P.~J.~E. {Peebles} and J.~T. {Yu}, {\em \apj} {\bf 162}  (1970)   815.

\bibitem{cobedmr}
E.~L. {Wright}, C.~L. {Bennett}, K.~{Gorski}, G.~{Hinshaw} and G.~F. {Smoot},
  {\em \apjl} {\bf 464}  (1996)   L21.

\bibitem{BOOMERANG}
{W. C. Jones et al.}, {\em \apj} {\bf 647}  (2006) 823.

\bibitem{acbar}
{C. L. Reichardt et al.}, {\em \apj} {\bf 694}  (2009) 1200.

\bibitem{wmap}
{D. Larson et al.}, {\em \apjs} {\bf 192}  (2011)  ~16.

\bibitem{spt}
{R. Keisler et al.}, {\em \apj} {\bf 743}  (2011)  ~28.

\bibitem{act}
{R. Hlozek et al.}, {\em \apj} {\bf 749}  (2012)  ~90.

\bibitem{d1975}
V.~K. {Dubrovich}, {\em Soviet Astronomy Letters} {\bf 1}  (1975)   196.

\bibitem{rcs2006}
J.~A. {Rubi{\~n}o-Mart{\'{\i}}n}, J.~{Chluba} and R.~A. {Sunyaev}, {\em \mnras}
  {\bf 371}  (2006) 1939.

\bibitem{kogut}
A.~{Kogut}, private communication  (2011).

\bibitem{planck}
{Planck Collaboration}, Planck blue book  (2005).

\bibitem{basu}
K.~{Basu}, C.~{Hern{\'a}ndez-Monteagudo} and R.~A. {Sunyaev}, {\em \aap} {\bf
  416}  (2004) 447.

\bibitem{cmbfast}
U.~{Seljak} and M.~{Zaldarriaga}, {\em \apj} {\bf 469}  (1996)   437.

\bibitem{dod}
S.~{Dodelson}, {\em {Modern cosmology}} (Modern cosmology / Scott
  Dodelson.~Amsterdam (Netherlands): Academic Press.~ISBN 0-12-219141-2, 2003).

\bibitem{wmap7}
{E. Komatsu et al.}, {\em \apjs} {\bf 192}  (2011)  ~18.

\bibitem{wmap3}
{G. Hinshaw et al.}, {\em \apjs} {\bf 148}  (2003) 63.

\bibitem{kk2003}
M.~{Kamionkowski} and L.~{Knox}, {\em \prd} {\bf 67}  (2003)   063001,
  \href{http://arxiv.org/abs/astro-ph/0210165}{{\ttfamily astro-ph/0210165}}.

\bibitem{cs2004}
J.~{Chluba} and R.~A. {Sunyaev}, {\em \aap} {\bf 424}  (2004) 389.

\bibitem{ks2013}
R.~A. {Sunyaev} and R.~{Khatri}, {\em \jcap} {\bf 3} (March 2013)  ~12,
  \href{http://arxiv.org/abs/1302.6571}{{\ttfamily arXiv:1302.6571
  [astro-ph.CO]}}.

\bibitem{zs1969}
Y.~B. {Zeldovich} and R.~A. {Sunyaev}, {\em \apss} {\bf 4}  (1969) 301.

\bibitem{sz1972}
R.~A. {Sunyaev} and Y.~B. {Zeldovich}, {\em Comments on Astrophysics and Space
  Physics} {\bf 4}  (1972)   173.

\bibitem{ks2012b}
R.~{Khatri} and R.~A. {Sunyaev}, {\em \jcap} {\bf 9}  (2012)  ~16.

\bibitem{k1956}
A.~S. {Kompaneets}, {\em Zh. Eksp. Teor. Fiz.} {\bf 31}  (1956) 876.

\bibitem{cs2011}
J.~{Chluba} and R.~A. {Sunyaev}, {\em \mnras} {\bf 419}  (2012) 1294.

\bibitem{sz1972b}
R.~A. {Sunyaev} and Y.~B. {Zeldovich}, {\em \aap} {\bf 20} (August 1972)   189.

\bibitem{co1999}
R.~{Cen} and J.~P. {Ostriker}, {\em \apj} {\bf 514}  (1999) 1.

\bibitem{co2006}
R.~{Cen} and J.~P. {Ostriker}, {\em \apj} {\bf 650}  (2006) 560.

\bibitem{ns2001}
B.~B. {Nath} and J.~{Silk}, {\em \mnras} {\bf 327}  (2001) L5.

\bibitem{actpol}
{M. D. Niemack et al.}, { {ACTPol: a polarization-sensitive receiver for the
  Atacama Cosmology Telescope}} {\em Society of Photo-Optical Instrumentation
  Engineers (SPIE) Conference Series} {\bf 7741} (2010), p. 77411.

\bibitem{sptpol}
{J. J. McMahon et al.}, { {SPTpol: an instrument for CMB polarization}}, in
  {\em American Institute of Physics Conference Series\/},  eds. B.~{Young},
  B.~{Cabrera} and A.~{Miller}, American Institute of Physics Conference
  Series, Vol.~1185 (2009), pp. 511--514.

\bibitem{ksc2012}
R.~{Khatri}, R.~A. {Sunyaev} and J.~{Chluba}, {\em \aap} {\bf 540}  (2012)
  A124.

\bibitem{wmapdipole}
{N. Jarosik et al.}, {\em \apjs} {\bf 192}  (2011)  ~14.

\bibitem{zis1972}
Y.~B. {Zeldovich}, A.~F. {Illarionov} and R.~A. {Sunyaev}, {\em Soviet Journal
  of Experimental and Theoretical Physics} {\bf 35}  (1972)   643.

\bibitem{cks2012}
J.~{Chluba}, R.~{Khatri} and R.~A. {Sunyaev}, {\em \mnras} {\bf 425}  (2012)
  1129.

\bibitem{zl1970}
Y.~B. {Zeldovich} and E.~V. {Levich}, {\em Soviet Journal of Experimental and
  Theoretical Physics Letters} {\bf 11}  (1970) 35.

\bibitem{ls1971}
E.~V. {Levich} and R.~A. {Sunyaev}, {\em \sovast} {\bf 15}  (1971)   363.

\bibitem{is1975b}
A.~F. {Illarionov} and R.~A. {Sunyaev}, {\em \sovast} {\bf 18}  (1975) 413.

\bibitem{sz1970}
R.~A. {Sunyaev} and Y.~B. {Zeldovich}, {\em \apss} {\bf 7}  (1970) 20.

\bibitem{ksvw2010}
J.~{Klaers}, J.~{Schmitt}, F.~{Vewinger} and M.~{Weitz}, {\em \nat} {\bf 468}
  (2010) 545.

\bibitem{dt1992}
S.~{Dodelson} and M.~S. {Turner}, {\em \prd} {\bf 46} (October 1992) 3372.

\bibitem{fdt1993}
B.~D. {Fields}, S.~{Dodelson} and M.~S. {Turner}, {\em \prd} {\bf 47} (May
  1993) 4309, \href{http://arxiv.org/abs/arXiv:astro-ph/9210007}{{\ttfamily
  arXiv:astro-ph/9210007}}.

\bibitem{dolgov}
A.~D. {Dolgov}, {\em \physrep} {\bf 370} (November 2002) 333,
  \href{http://arxiv.org/abs/arXiv:hep-ph/0202122}{{\ttfamily
  arXiv:hep-ph/0202122}}.

\bibitem{mm2005}
G.~{Mangano}, G.~{Miele}, S.~{Pastor}, T.~{Pinto}, O.~{Pisanti} and P.~D.
  {Serpico}, {\em Nuclear Physics B} {\bf 729}  (2005) 221.

\bibitem{kawano}
L.~Kawano, { {Let's Go: Early Universe. Guide to Primordial Nucleosynthesis
  Programming,}}
  {\url{http://www-thphys.physics.ox.ac.uk/people/SubirSarkar/bbn.html}},
  (1988).

\bibitem{wagoner}
R.~V. {Wagoner}, {\em \apj} {\bf 179} (January 1973) 343.

\bibitem{lightman}
A.~P. {Lightman}, {\em \apj} {\bf 244}  (1981) 392.

\bibitem{thorne81}
K.~S. {Thorne}, {\em \mnras} {\bf 194}  (1981) 439.

\bibitem{sz1970d}
R.~A. {Sunyaev} and Y.~B. {Zeldovich}, {\em Comments on Astrophysics and Space
  Physics} {\bf 2} (March 1970)  ~66.

\bibitem{dd1982}
L.~{Danese} and G.~{de Zotti}, {\em \aap} {\bf 107}  (1982) 39.

\bibitem{ks2012}
R.~{Khatri} and R.~A. {Sunyaev}, {\em \jcap} {\bf 6}  (2012)  ~38.

\bibitem{lifshitz}
E.~M. {Lifshitz}, {\em {J. Phys. (USSR)}} {\bf 10}  (1946)   116.

\bibitem{ly2006}
P.~{McDonald}, U.~{Seljak}, S.~{Burles}, D.~J. {Schlegel}, D.~H. {Weinberg},
  R.~{Cen}, D.~{Shih}, J.~{Schaye}, D.~P. {Schneider}, N.~A. {Bahcall}, J.~W.
  {Briggs}, J.~{Brinkmann}, R.~J. {Brunner}, M.~{Fukugita}, J.~E. {Gunn}, {\v
  Z}.~{Ivezi{\'c}}, S.~{Kent}, R.~H. {Lupton} and D.~E. {Vanden Berk}, {\em
  \apjs} {\bf 163}  (2006) 80.

\bibitem{ssm2006}
U.~{Seljak}, A.~{Slosar} and P.~{McDonald}, {\em \jcap} {\bf 10}  (2006)  ~14.

\bibitem{silk}
J.~{Silk}, {\em ApJ} {\bf 151}  (1968)   459.

\bibitem{kaiser}
N.~{Kaiser}, {\em MNRAS} {\bf 202}  (1983) 1169.

\bibitem{sz1970b}
R.~A. {Sunyaev} and Y.~B. {Zeldovich}, {\em Ap\&SS} {\bf 9}  (1970) 368.

\bibitem{daly1991}
R.~A. {Daly}, {\em \apj} {\bf 371}  (1991) 14.

\bibitem{hss94}
W.~{Hu}, D.~{Scott} and J.~{Silk}, {\em ApJl} {\bf 430}  (1994) L5.

\bibitem{ksc2012b}
R.~{Khatri}, R.~A. {Sunyaev} and J.~{Chluba}, {\em \aap} {\bf 543}  (2012)
  A136.

\bibitem{pz2012}
E.~{Pajer} and M.~{Zaldarriaga}, {\em ArXiv e-prints}   (2012)
  \href{http://arxiv.org/abs/1206.4479}{{\ttfamily arXiv:1206.4479}}.

\bibitem{dent2012}
J.~B. {Dent}, D.~A. {Easson} and H.~{Tashiro}, {\em \prd} {\bf 86}  (2012)
  023514.

\bibitem{ceb2012}
J.~{Chluba}, A.~L. {Erickcek} and I.~{Ben-Dayan}, {\em \apj} {\bf 758}  (2012)
  ~76.

\bibitem{pajer2012}
E.~{Pajer} and M.~{Zaldarriaga}, {\em Physical Review Letters} {\bf 109}
  (2012)   021302.

\bibitem{ganc2012}
J.~{Ganc} and E.~{Komatsu}, {\em \prd} {\bf 86}  (2012)   023518.

\bibitem{pb2009}
P.~{Procopio} and C.~{Burigana}, {\em \aap} {\bf 507}  (2009) 1243.

\bibitem{fengreview}
J.~L. {Feng}, {\em \araa} {\bf 48}  (2010) 495.

\bibitem{fengsusy}
J.~L. Feng, A.~Rajaraman and F.~Takayama, {\em Phys. Rev. D} {\bf 68}  (2003)
  063504.

\bibitem{fengkk}
J.~L. Feng, A.~Rajaraman and F.~Takayama, {\em Phys. Rev. D} {\bf 68}  (2003)
  085018.

\bibitem{csbook}
A.~{Vilenkin} and E.~P.~S. {Shellard}, {\em {Cosmic Strings and Other
  Topological Defects}} (Cambridge University Press, Cambridge, 2000).

\bibitem{vilenkin1988}
A.~{Vilenkin}, {\em \nat} {\bf 332}  (1988)   610.

\bibitem{tsv2012}
H.~{Tashiro}, E.~{Sabancilar} and T.~{Vachaspati}, {\em \prd} {\bf 85}  (2012)
   103522.

\bibitem{mrs2005}
A.~{Mirizzi}, G.~G. {Raffelt} and P.~D. {Serpico}, {\em \prd} {\bf 72}  (2005)
   023501.

\bibitem{chrr2011}
D.~{Cadamuro}, S.~{Hannestad}, G.~{Raffelt} and J.~{Redondo}, {\em \jcap} {\bf
  2}  (2011)  ~3.

\bibitem{ellis2013}
G.~F.~R. {Ellis}, R.~{Poltis}, J.-P. {Uzan} and A.~{Weltman}, {\em ArXiv
  e-prints}   (2013) \href{http://arxiv.org/abs/1301.1312}{{\ttfamily
  arXiv:1301.1312 [astro-ph.CO]}}.

\bibitem{jko2000}
K.~{Jedamzik}, V.~{Katalini{\'c}} and A.~V. {Olinto}, {\em Physical Review
  Letters} {\bf 85}  (2000) 700.

\bibitem{lochan2012}
K.~{Lochan}, S.~{Das} and A.~{Bassi}, {\em \prd} {\bf 86}  (2012)   065016.

\bibitem{tn04}
H.~Tashiro and N.~Sugiyama, {\em Phys. Rev. D} {\bf 78}  (2008)   023004.

\end{thebibliography}
\end{document}